\documentclass[conference, a4paper]{IEEEtran}

\IEEEoverridecommandlockouts

\usepackage{cite}
\usepackage{amsmath,amssymb,amsfonts}
\usepackage{algorithmic}
\usepackage{graphicx}
\usepackage{textcomp}
\usepackage{xcolor}

\usepackage{siunitx}
\usepackage{hyperref}
\usepackage{url}

\newcommand \IOC {I4oC}
\newcommand \EOC {\textrm{EoC}}
\newcommand \TETH {\textit{ETH}}
\newcommand \TCAN {\textit{CAN}}

\def\BibTeX{{\rm B\kern-.05em{\sc i\kern-.025em b}\kern-.08em
    T\kern-.1667em\lower.7ex\hbox{E}\kern-.125emX}}

\usepackage{lipsum}
\usepackage[pscoord]{eso-pic}
\newcommand{\placetextbox}[3]{
  \setbox0=\hbox{#3}
  \AddToShipoutPictureFG*{
    \put(\LenToUnit{#1\paperwidth},\LenToUnit{#2\paperheight}){\vtop{{\null}\makebox[0pt][c]{#3}}}%
  }%
}%

\title{Composite CAN XL-Ethernet Networks for 
Next-Gen Automotive and Automation Systems\\
}

\author{
    \IEEEauthorblockN{
    Gianluca Cena, 
    Stefano Scanzio, 
    and Adriano Valenzano 
    }
    \IEEEauthorblockA{
    National Research Council of Italy (CNR--IEIIT), Italy.\\ 
    Email: 
    \{gianluca.cena, stefano.scanzio, adriano.valenzano\}@cnr.it
    }
}

\begin{document}
\placetextbox{0.5}{1}{This is the author's version of an article that has been published in this journal.}
\placetextbox{0.5}{0.985}{Changes were made to this version by the publisher prior to publication.}
\placetextbox{0.5}{0.97}{The final version of record is available at \href{https://doi.org/10.1109/WFCS57264.2023.10144116}{https://doi.org/10.1109/WFCS57264.2023.10144116}}%
\placetextbox{0.5}{0.05}{Copyright (c) 2023 IEEE. Personal use is permitted.}
\placetextbox{0.5}{0.035}{For any other purposes, permission must be obtained from the IEEE by emailing pubs-permissions@ieee.org.}%

\maketitle

\thispagestyle{empty}
\pagestyle{empty}

\begin{abstract}
New generation electrified and self-driving vehicles require much higher performance and flexibility for onboard digital communications than Controller Area Networks may offer. For this reason, automotive Ethernet is often regarded as the next de facto standard technology in these contexts, and by extension for networked embedded systems as well. However, an abrupt and drastic move from CAN to Ethernet is likely to cause further cost increases, which can be hardly tolerated by buyers.

This paper analyzes the third generation of CAN, termed CAN XL, and studies how interoperability can be ensured with Ethernet. Likely, composite CAN XL-Ethernet networks are the key for getting the best of both worlds, not only in the automotive domain but also for sensing and control in scenarios like building automation, wired sensor networks, and low-cost networked embedded systems with real-time constraints.
\end{abstract}


\section{Introduction}
Controller Area Network (CAN) \cite{ISO11898} 
has been the solution of choice for onboard digital communications in the automotive industry since the early 1990s.
Besides cars, it is widely adopted also in trucks, buses, and heavy-duty vehicles, as well as in a multiplicity of networked embedded systems, including automatic dispensers, elevators, and medical equipment, to cite a few.
The two main limitations of classic CAN concern the transmission speed, 
which is not allowed to exceed $\SI{1}{Mb/s}$ because of its peculiar medium access technique 
(bit rate is often limited to $\SI{500}{kb/s}$), and payload, which can include up to $\SI{8}{B}$ at most.
A first attempt to overcome above limitations was made in 2012 with the proposal Controller Area Network with Flexible Data-Rate (CAN FD) \cite{CANFD}, now included in the ISO 11898-1:2015 specification.
By allowing different speeds for the arbitration and data transmission phases, 
the bit rate can be increased up to $\SI{8}{Mb/s}$ when signal improvement capability (SIC) transceivers are used.
Moreover, the maximum payload size grows up to $\SI{64}{B}$,
which is enough to encode real-time process data in most of the existing control systems.

In spite of the tangible performance boost over classic CAN,
CAN FD will be hardly able to face the competition of next-generation automotive networks in the long term, 
which likely will rely on the proven and ubiquitous Ethernet technology.
Moving to Ethernet opens a whole range of possibilities to carmakers: 
besides pure transmission speed, an unprecedented level of synergy is potentially enabled among different application scenarios.
As a matter of fact, wiring in home-office contexts (and, more recently, also in the industrial ones) now relies for the most part on this technology.
On the other hand, the installed CAN base in the automotive domain 
(and the related know-how acquired by designers in the past three decades) are undeniably huge, 
too big to be simply thrown away with no severe economic consequences.

To narrow the gap between CAN and Ethernet, a new proposal is currently being defined, known as CAN eXtra Long (CAN XL),
which includes modifications to both the physical (PHY) \cite{DS610-3} and data-link (DL) \cite{DS610-1} layers 
and enables speeds well above CAN FD.
Moreover, its much larger payload makes it easy to transfer complex/structured information including, 
e.g., video streams produced by front and rear cameras.
Finally, CAN XL features high versatility, and permits to encapsulate IEEE 802.3 MAC frames thanks to its multiplexing/demultiplexing functions.
This implies that it is capable to emulate an Ethernet interface to the upper protocol layers, 
which applications can exploit to communicate with the same paradigms available in home/office/industrial scenarios.
In particular, layering the conventional TCP/UDP/IP protocol stack above CAN XL offers the users 
the ability to re-use high-level IETF standard protocols (HTTP, MQTT, OPC UA, RTP, etc.) 
and the related implementations (both commercial and open source).

The above approach permits Ethernet and CAN XL to seamlessly coexist in the same distributed system,
which is of utmost importance in the automotive field.
Unlike solutions based on pure Ethernet, which have to be re-conceived from scratch,
composite CAN XL-Ethernet 
systems
provide an effortless migration path from the existing designs to future ones.
Besides, the ability to easily connect CAN XL sub-networks to the Ethernet backbone can be also beneficial for deploying CAN-based sensor and actuator networks (CSAN) in industrial plants.
They resemble wireless sensor networks (WSN), but are faster, much more reliable, 
and can possibly power devices on the same cable used for communication.

In this paper some non-custom approaches to make Ethernet and CAN fully interoperable are presented, 
and their advantages and drawbacks are discussed.
We only consider onboard communications among ECUs, as vehicle-to-infrastructure (V2I) and vehicle-to-vehicle (V2V) communications rely on other technologies, e.g., 4G/5G networks \cite{V2X17}.
The paper is structured as follows: 
Section~\ref{sec:canxl} briefly reviews the most important features of CAN XL,
whereas encapsulation of Ethernet and IP are considered in Sections~\ref{sec:EoC} and \ref{sec:IoC}, respectively.
Section~\ref{sec:CANbridge} considers additional CAN-to-CAN store and forward functions, while Section~\ref{sec:Conc} draws some conclusions.

\section{CAN XL Data-Link Layer}
\label{sec:canxl}
The data-link layer of CAN XL is structured in two tiers, Medium Access Control (MAC) and Logical Link Control (LLC).
The former is now almost stable and waiting for ISO standardization.
Instead, definition of the latter is underway, therefore several aspects are still undecided and may change.

\subsection{Medium Access Control}
The CAN XL MAC is not dissimilar from classic CAN and CAN FD, which ensures good backward compatibility.
Access to the shared transmission support (CAN bus) relies on deterministic contention
(\textit{bitwise} arbitration) carried out on the priority field, located just after the Start of Frame (SOF) bit.
This enables fine-grained traffic \textit{prioritization}, which permits to check 
(through schedulability analysis) whether or not real-time constraints are met for data exchanges \cite{davisControllerAreaNetwork2007}.
Unlike previous CAN versions, only $\SI{11}{b}$ identifiers are supported:
the extended $\SI{29}{b}$ format has been dropped in CAN XL, 
since other means are made available to enlarge its addressing capabilities.
This is not a real limitation, since the identifier extension is rarely exploited for arbitration, but serves to convey additional information that can not fit in the payload.

Similarly to CAN FD, frame exchange in CAN XL distinguishes between arbitration and data phase.
However, behavior of the transceiver in CAN XL is changed in the data phase (FAST TX Mode) so that there are no longer a dominant and a recessive level, but two symmetric values ($0$ and $1$) as in the other kinds of networks.
The arbitration to data sequence (ADS) and data to arbitration sequence (DAS) fields surround the part of frame sent at higher speed, and were designed to ensure a graceful transition between phases.
This enables tangibly higher bit rates, which at present can be as high as $\SI{20}{Mb/s}$ and, in theory, can be raised further.

The most direct competitor of CAN XL when bit rates in the order of $\SI{10}{Mb/s}$ are considered is multidrop 10BASE-T1S with PHY-Level Collision Avoidance (PLCA) in IEEE 802.3cg \cite{IEEE8023cg}, 
defined as part of automotive Ethernet.
PLCA loosely resembles Byteflight and its \textit{linear} arbitration scheme, but has been conceived from scratch to comply to Ethernet and ensure a high degree of backward compatibility.
A Reconciliation Sublayer (RS) is foreseen between the Ethernet MAC and PHY that provides stations fair round-robin access to the bus.
A master node, characterized by identification number (ID) $0$, 
is in charge of sending beacons that synchronize operations of the other nodes.
Following the beacon, every PLCA node is provided a single transmission opportunity, in increasing ID number (from $0$ to $N$).
Nodes with no buffered frames waiting for transmission are skipped after a very short \textit{TO\_TIMER} (typically \SI{20}{b}),
which makes this mechanism very efficient.
The waiting time before a PLCA node can send a queued frame is bounded, and latency can be easily calculated.

It is generally agreed that bitwise arbitration is superior to linear arbitration  \cite{cenaPropertiesFlexibleTime2006}, especially when message streams have different deadlines.
Conversely, PLCA shows fairer behavior when data exchanges are characterized by similar timings.
However, a slight modification to the MAC \cite{cenaFairAccessMechanism2022},
which defines two distinct durations for the intermission between consecutive frames,
enables fair round-robin access in CAN as well, besides prioritized access.
Moreover, in CAN XL there is no need to have a master node, which improves robustness.

Another relevant improvement of CAN XL over the previous versions concerns the maximum size of the data field, 
which has been increased up to $\SI{2048}{B}$.
This is strictly larger than Ethernet frames, which can be thus embedded into CAN XL frames. 
This larger data field requires a stronger frame check sequence (FCS) than classic CAN and CAN FD, implemented as a suitable CRC-32.
It is worth noting that in CAN XL a second CRC-13 (preface CRC) is foreseen that covers (part of) the frame header, which improves robustness further by providing for it additional protection.

\subsection{Logical Link Control}
The CAN XL LLC is much richer than its former versions, and is planned to include a number of functions for supporting higher-layer protocols.
Below is a list of \textit{basic} LLC functions, whose behavior is already stable, which are encoded in the frame's header (control field):
\begin{itemize}
\item \textit{Multiplexing}:
As for Ethernet, a number of protocols can be easily layered atop CAN XL.
They are distinguished thanks to the service data unit type (SDT) field, hereinafter also referred to as SDU type, 
which consists in one octet and tells receivers how the following fields have to be interpreted. 
SDT resembles the EtherType (ET) field in Ethernet, and is mainly used for 
transporting
frames belonging to other protocols, including Ethernet, classic CAN, and CAN FD.
Allowed SDT values are defined in companion standards.

\item \textit{Virtualization}:
This function resembles virtual local area networks (VLAN) as defined in IEEE 802.1Q 
\cite{IEEE8021Q},
and can be used to partition the physical support into logically independent virtual networks.
In real-time networks like CAN, interdependence between virtual networks defined on the same communication support and due to the interference at the MAC layer is unavoidable.
This means that feasibility analysis can not be carried out separately for them, hence impacting on composability, which must be supported by other means.
Virtualization can be advantageous in networks interconnected by bridges, 
like composite CAN XL-Ethernet ones, and could be exploited for security purposes as well.

\item \textit{Filtering}:
To unburden the microcontroller in receivers from the task of filtering out (in software) messages not targeted to them, 
the acceptance field (AF) has been included in the frame header. 
AF consists in four octets and enables message filtering to be performed in hardware by the CAN controller.
Its usage depends on the specific upper layer protocol, as defined by SDT.
This enhances the behavior of existing controllers, which typically perform filtering on the identifier field only.
\end{itemize}

A number of \textit{extended} LLC functions are also envisaged,
which are typically enabled by setting the simple extended content (SEC) flag in the control field of the MAC header.
In this case, the relevant headers for these functions are inserted at the beginning of the data field.
\begin{itemize}
\item \textit{Fragmentation}:
The size of the data field in CAN XL is enough to accommodate also the largest Ethernet frame.
However, a larger payload implies a longer duration for transmissions on the bus, which in turn increases blocking time and interference,
negatively affecting responsiveness.
For example, a full size CAN XL frame (including $\SI{2048}{B}$ in the data field) sent with arbitration bit rate equal to $\SI{500}{kb/s}$ and data bit rate set to $\SI{16}{Mb/s}$ lasts about $\SI{1.20}{ms}$,
which is six times higher than the worst-case blocking time in a CAN network with the same nominal speed
(about $\SI{0.22}{ms}$ for $\SI{11}{b}$ identifiers).
Thus, embedding Ethernet frames whose size equals the Maximum Transmission Unit (MTU=$\SI{1500}{B}$) impacts on feasibility analysis \cite{davisControllerAreaNetwork2007},
and is likely to make 
real-time
messages exceed their intended deadlines.
To solve this issue, CAN XL is planned to include a standard fragmentation protocol in the LLC, which loosely resembles ISO-TP \cite{ISO15765}.

\item \textit{Security}:
Advanced driver-assistance systems (ADAS) 
may incur in serious security issues.
In fact, any security breach to X-by-wire systems (X may stand for throttle, brake, shift, steer, etc.) 
may turn the vehicle into a weapon and jeopardize the safety of both passengers and nearby people and objects.
For this reason, data-link layer security measures similar to IEEE 802.1AE (MACsec) \cite{MACSEC},
that is, authentication and encryption,
are planned for CAN XL.

\item \textit{Aggregation}:
This function may be helpful when small data packages are generated by applications.
In this case, gathering a number of them in the same CAN XL frame may improve communication efficiency.

\end{itemize}

\begin{figure*}[h]%[htbp]
    \centerline{\includegraphics[width=0.95 \linewidth]{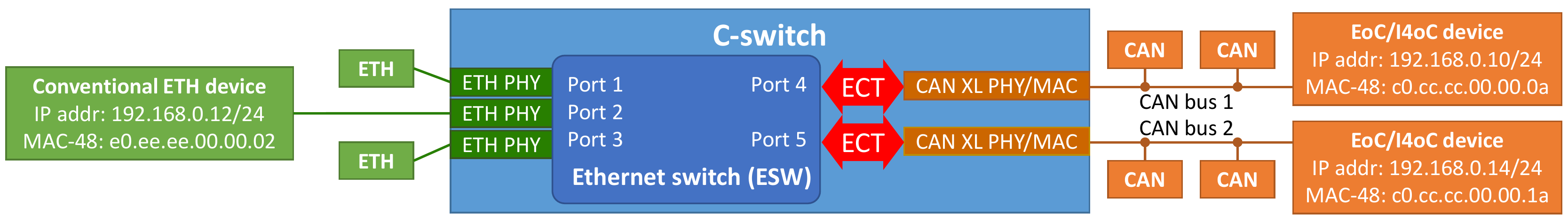}}
    \caption{C-switch internal architecture: a conventional Ethernet switch (ESW) is embedded 
    that performs backward learning and store\&forward operations.}
    \label{fig:CSWITCH}
\end{figure*}

\section{Ethernet over CAN XL}
\label{sec:EoC}
The simplest way to make devices provided with Ethernet and CAN interfaces communicate with each other 
is by encapsulating Ethernet frames inside CAN XL frames.
This feature, which has been explicitly sought since CAN XL inception, will be referred to as Ethernet over CAN XL (\EOC).
It is completely transparent to Ethernet devices, which see \EOC\ devices connected to a CAN bus as if they were conventional Ethernet stations.
A second alternative, not considered here, is to transport CAN frames in Ethernet (CoE), as done, e.g., in \cite{SIES15-C2E}.
In this case, several options are possible, including the adoption of time-sensitive networking (TSN)
to preserve timings across the whole composite network \cite{ICC17-Zervas}.

\subsection{C-switch}
In the composite networks we are taking into account, end-devices are typically provided with a single port, of either \TETH\ or \TCAN\ type.
Since these technologies differ at both the PHY and MAC layers,
the related devices cannot be interconnected by exploiting pure-software solutions.
Conversely, specific intermediate network equipment is needed that operates according to the store and forward principle.
It takes care of translating frames belonging to the two networks and accommodates their different speeds.
In the following such a device will be called \textit{composite switch} (\mbox{C-switch}).

A \mbox{C-switch} is a multiport bridge where ports can be of either \TETH\ or \TCAN\ type.
Further distinctions can be made for either kind of port.
In the \TETH\ case, other options are available besides RJ-45 connectors, like Single Pair Ethernet (SPE).
Concerning \TCAN\ one can distinguish among classic CAN, CAN FD, and CAN XL.
CAN version is not irrelevant: sending a CAN FD/XL frame on a bus to which legacy nodes are attached causes repeated failures that prevent communication \cite{ICC15-Adam}.
These aspects are not considered in the following, as we mainly focus on store and forward operations.
Clearly, every end-device can be connected only to the ports on the \mbox{C-switch} of corresponding type.
Unlike Ethernet switches used in home-office and industrial networks, physical connections are not only point-to-point,
since both automotive Ethernet and CAN support multidrop bus topologies.
Wireless extensions are also possible:
Wi-Fi closely resembles Ethernet, as both share the same address space (MAC-48), 
similar MAC techniques (nodes access the shared medium without requiring prior permission), and similarly-sized MTUs.
Practically, an access point can be included in the \mbox{C-switch} that permits CAN nodes to be accessed over the air.

\mbox{C-switch} operation is straightforward and resembles conventional Ethernet switches.
The main difference is that ports of \TCAN\ type require some means to encapsulate/decapsulate Ethernet into/from CAN XL.
Practically, its internal architecture can be schematically described as shown in Fig.~\ref{fig:CSWITCH}.
At the core of a \mbox{C-switch}, a conventional Ethernet switch (ESW) is found that performs selective forwarding
(no need to re-invent the wheel).
Ports of the \mbox{C-switch} of type \TETH\ are directly connected to the corresponding EWS ports,
whereas ports of type \TCAN\ are connected through an \EOC\ tunneller (ECT), 
depicted as a thick red arrow and implemented as a piece of code that performs encapsulation and decapsulation.
In the following, the term ``\EOC\ frame'' denotes a CAN XL frame that embeds an Ethernet frame according to the rules stated in \cite{DS611-1}, which include setting SDT to the ``IEEE 802.3'' value.

Whenever an \EOC\ frame is received on a port of type \TCAN\ of the \mbox{C-switch},
the embedded Ethernet frame is extracted by the ECT and transferred to the associated ESW port.
No additional filtering is carried out by ECT on ingress \EOC\ frames, e.g., based on AF,
as selective forwarding is performed by ESW.
In the opposite direction, when an Ethernet frame is forwarded by ESW on a port of type \TCAN\ of the \mbox{C-switch}, 
ECT embeds such frame in an \EOC\ frame, which is then sent on the associated CAN bus.
The priority field of outgoing CAN XL frames shall be suitably selected to preserve timeliness constraints, 
preventing at the same time any clashes with other CAN nodes connected to that bus.
In fact, CAN is unable to resolve contentions among concurrent frame exchanges when nodes are using the same priority.
Defining uniform and agreed rules for assigning priorities in CAN XL is still an open question, and is left for future works.

\subsection{\EOC\ Devices}

Let us denote as ``\EOC\ nodes'' those end-devices that are provided with an interface of type \TCAN\ and all the necessary means to send and receive \EOC\ frames.
In practice, they include a thin protocol layer, located above the CAN XL data-link layer, that emulates Ethernet in software by exploiting 
the SDU type.
The primary purpose of this entity is to offer a suitable interface to the upper protocol layers (e.g., the conventional TCP/UDP/IP protocol suite),
which are enabled to run with no significant changes, hence easing porting operations.
A socket-based interface can be additionally provided to applications, which resembles raw (Ethernet) sockets.
It is worth noting that sockets are often envisaged also for conventional CAN data exchanges, 
e.g., \cite{cenaSocketInterfaceCAN2007} or, more recently, SocketCAN \cite{ICC12-Hart, ICC15-Hart}.
In this architecture, \EOC\ nodes are characterized by a double protocol stack, which enables contextual CAN-Ethernet operations.
For instance, real-time CAN FD messages can be exchanged with other CAN devices connected to the same bus
and, at the same time, a TCP connection is established  for communicating with Ethernet devices via the \mbox{C-switch}.

Addressing in \EOC\ nodes relies on the MAC-48 destination and source address fields (SA and DA) in the embedded Ethernet frame.
This means that every \EOC\ node must be assigned a unique MAC-48 address, as for the Ethernet ones.
To unburden nodes from the need of checking (in software) the DA field for every received \EOC\ frame,
part of this field (four out of six octets) is copied in the AF field by the sender.
As a consequence, filtering on AF may lead to clashes (false positives), even though such events are statistically very rare in real networks.
In the case a clash occurs, a subsequent check performed by the microcontroller on the embedded DA field permits to break ties.
Since an \EOC\ node must accept both the frames with an individual DA targeted to it
and those with a group DA (broadcast and multicast),
the individual/group (I/G) bit of DA is included in AF.
The above kind of filtering is conceived to be performed in hardware by the CAN XL controller,
thus the amount of interrupts generated to the microcontroller is lowered to the bare minimum.

\begin{figure*}[htbp]
    \centerline{\includegraphics[width=0.89 \linewidth]{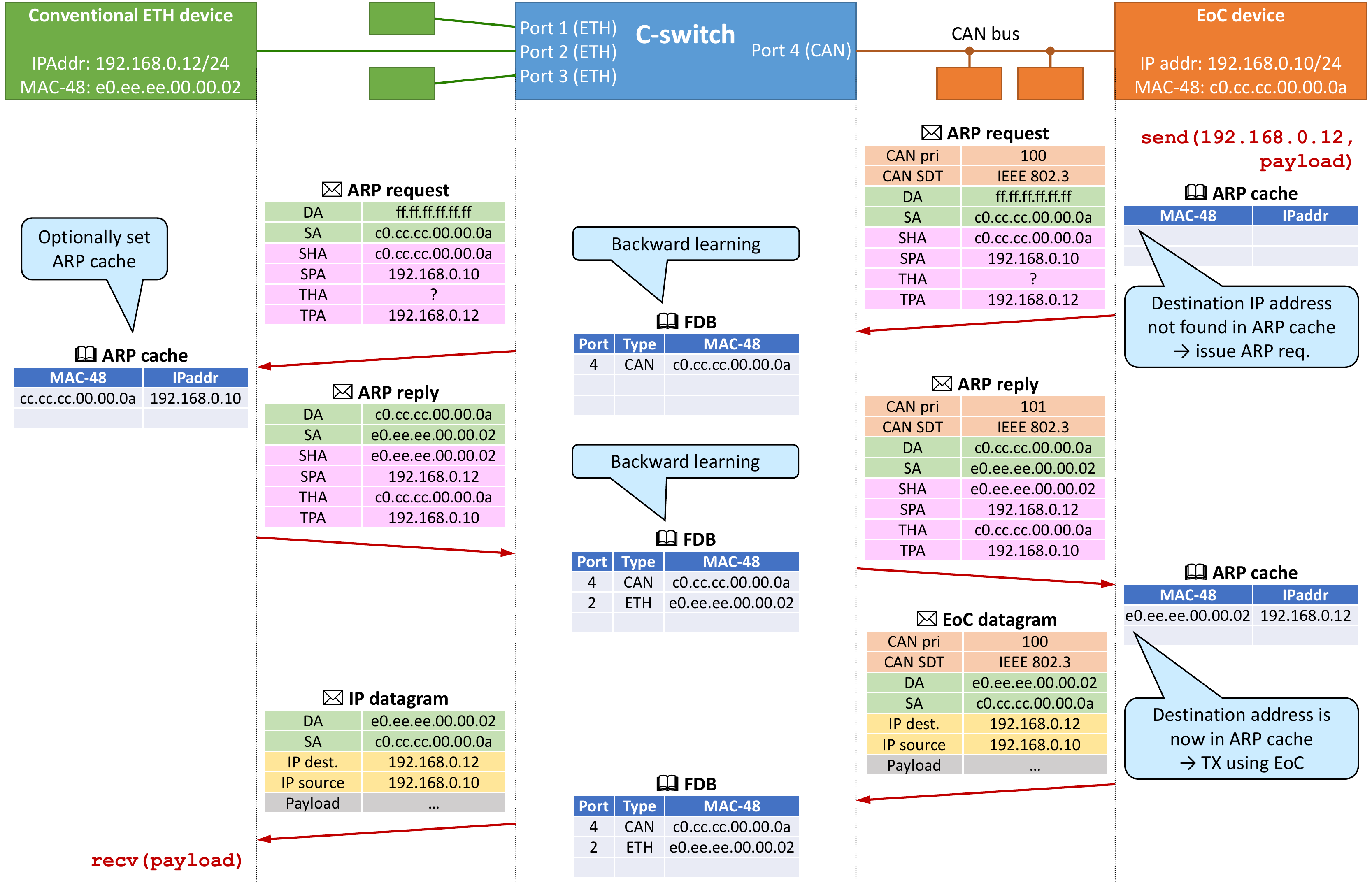}}
    \caption{Sample CAN-to-Ethernet datagram transmission using \EOC\ (baseline, operates exactly as switched Ethernet, besides ECT encapsulation).}
    \vspace{-0.1cm}
    \label{fig:CAN0}
    \vspace{-0.1cm}
\end{figure*}

\subsection{Selective Forwarding and Backward Learning}
The same backward learning technique used by switched Ethernet \cite{IEEE8021Q},
which relies on the SA field of the received frames to determine on which port nodes are connected,
is also exploited by the \mbox{C-switch} for \EOC\ frames.
In particular, a filtering database (FDB) has to be foreseen that,
when a frame is received on a given port, creates/updates an entry $\langle$SA,port$\rangle$.
This operation is performed by ESW, without requiring any modifications.
Since coding efforts almost completely concern ECT blocks, implementation costs are kept low.

Flooding of frames whose DA field is unknown to the switch, as well as forwarding of frames with group DA, 
is performed very efficiently on ports of type \TCAN\ by exploiting the multidrop nature of CAN busses.
Thus, it causes little additional traffic when CAN XL is employed to deploy sub-networks to which electronic control units (ECU) or, in embedded systems, sensors and actuators, are directly connected.

\subsection{Layering IP above \EOC}
Most of the times Ethernet is used to transport IP datagrams.
This is the case of Internet-enabled distributed applications, but applies almost identically to intranets.
In these contexts, mapping the addresses used at the network layer onto the related MAC addresses is an essential function.
To this extent, the address resolution protocol (ARP) has been defined \cite{rfc826},
which is customarily used (behind the scenes) by IP to translate its addresses in the corresponding MAC-48 addresses.
As depicted in Fig.~\ref{fig:CAN0}, the simplest solution to implement ARP in composite CAN XL-Ethernet networks is to map ARP messages on Ethernet frames, which in turn are encapsulated in CAN XL.
This is not the only possible approach. In fact, a new SDU type ``ARP'' could be in theory defined: ARP is a quite generic protocol, and suits a variety of network technologies.
We do not pursue this solution as ARP traffic is comparatively low due to caching.

ARP requests propagate over the entire composite network, which in our case is an embedded intranet made up of an arbitrary number of Ethernet subnetworks (that rely on either switches or SPE multidrop busses) and CAN segments, 
interconnected by \mbox{C-switches} to form any possible topology. %, including complex ones.
Clearly, the proposed solution shall work correctly also in the (unlikely but possible) cases where, in their travel from the source to the destination, frames are converted multiple times back and forth between Ethernet and CAN XL.
This means that, in the composite network, the spanning tree protocol (STP or RST) must be employed to prevent loops (they appear when, e.g., link redundancy is exploited to improve fault-tolerance).
In the solution envisaged above, loop detection and removal is carried out by \mbox{C-switches}. 
All it is needed is to encapsulate bridge protocol data units (BPDU) using \EOC,
in which case the spanning tree is constructed directly by ESWs, 
maintaining implementations costs low.

\section{IP over CAN}
\label{sec:IoC}
Modern in-vehicle systems foresee that part of the ECUs, e.g., devices involved in ADAS,
are connected through Ethernet and communicate using the TCP/UDP/IP protocol suite.
This is not the case of factory automation systems that rely on industrial Ethernet protocols like PROFINET, EtherCAT, and SERCOS III,  which are encapsulated in Ethernet using specific EtherType values.
For above embedded intranets a more efficient solution than \EOC\ can be devised,
we call IP over CAN (IoC), that maps IP directly on CAN.
A preliminary version of this approach was proposed in \cite{CENA20161}, but at that time CAN XL was not available yet.
For this reason a new proposal is made here that benefits from the higher performance and advanced features provided by the upcoming CAN version.

\subsection{Addressing and Encoding in \IOC}
Two non-negligible issues are encountered when IP datagrams are conveyed on classic CAN.
First, CAN relies on object addressing, whereas IP assumes the underlying network to be based on node addressing (i.e., to use a source-destination address pair).
Moreover, the base $\SI{11}{b}$ CAN identifier field is too short to properly support IP, unless the network is very small, hence extended $\SI{29}{b}$ identifiers (not available in CAN XL) should be used.
Second, the limited payload size makes fragmentation unavoidable (even for UDP/IP, \SI{28}{B} are required for headers), which implies low goodput.
These are no longer issues in CAN XL, as \EOC\ shows.
To improve communication efficiency further, two new SDU types can be defined for IPv4 and IPv6.
In the following we will only focus on the former (\IOC), which is likely the most relevant in the automotive field.
Extension to IPv6 could exploit, e.g., some of the solutions conceived for header compression in IPv6 over Low-Power Wireless Personal Area Networks (6LoWPAN).

A basic requirement with IP is that every interface of the hosts in the embedded network must be assigned a distinct IP address.
This is the same as Ethernet,
the main difference being that uniqueness of MAC-48 addresses is ensured by the centralized assignment of Organizationally Unique Identifiers (OUI), whereas for IP it is left to the system designer.
Embedded intranets exploit private addresses, which permit address assignment to be replicated identically for all instances of a given vehicle model.
High dependability and determinism are usually demanded in automotive and embedded systems, 
so static assignment is preferable.
However, non-critical functions (e.g., to temporarily connect through Wi-Fi a diagnostic/configuration tool, implemented as an app running on a smartphone/notebook) can rely on DHCP.

\begin{figure}%[htbp]
    \centerline{\includegraphics[width=1.0 \linewidth]{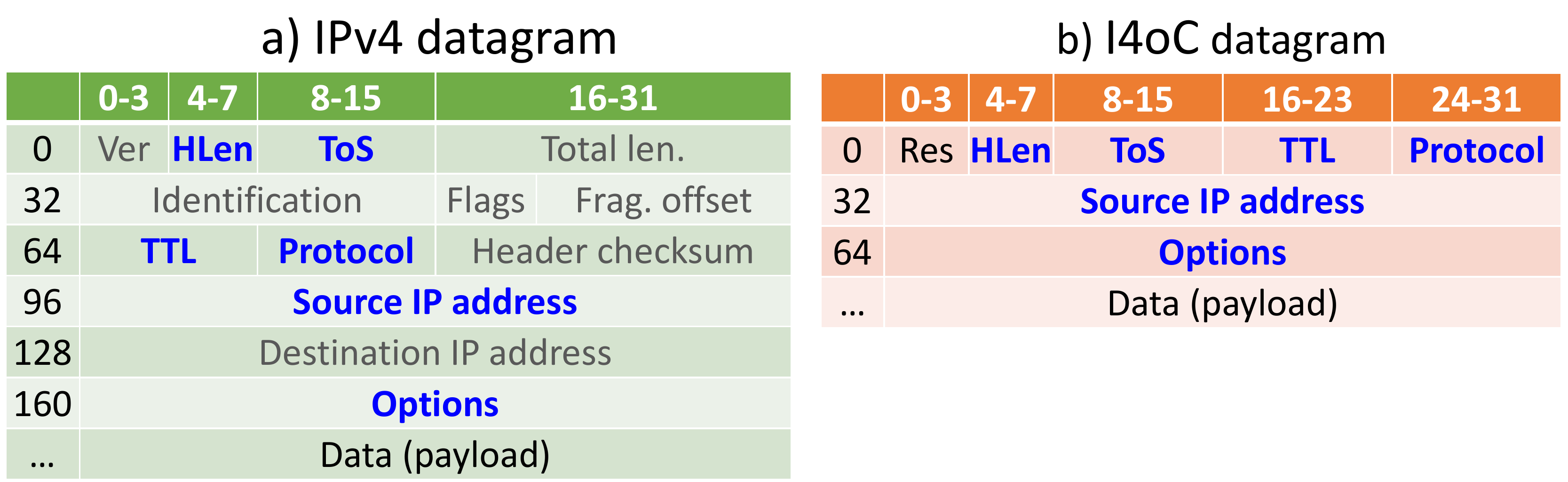}}
    \caption{Compact format of IP datagrams in \IOC.}
    \vspace{-0.2cm}
    \label{fig:IP}
    \vspace{-0.2cm}
\end{figure}

Embedding IP datagrams directly in CAN XL permits to save $\SI{14}{B}$ per datagram (Ethernet DA, SA, and ET fields).
Moreover, a more compact representation can be used for the IP header that saves another $\SI{12}{B}$, as shown in Fig.~\ref{fig:IP}.
Since the destination IP address is encoded in the AF, it can be omitted.
The same happens to the total length field (that can be provided by the MAC),
the header checksum (error detection offered by the double CRC in CAN XL is adequate),
and all fields related to fragmentation (when datagrams that exceed the maximal CAN XL payload are sent that require source-side fragmentation, \EOC\ is employed in the place of \IOC).
Overall, \IOC\ encoding is $\SI{26}{B}$ smaller than \EOC.

Saving is negligible when the payload is large (datagrams coming from Ethernet can include up to $\SI{1480}{B}$),
but it may be significant for small datagrams, 
e.g., process data sent using RTP and MQTT PUBLISH messages, or inquired through HTTP GET requests.
For example, if the datagram size is $\SI{64}{B}$,
transmission on Ethernet takes $\SI{90}{B}$ (including preamble, header, and FCS),
corresponding to $\SI{72}{\mu s}$ at $\SI{10}{Mb/s}$.
Sending the same datagram as \EOC\ on CAN XL with $\SI{500}{kb/s}$ nominal bit rate and
$\SI{16}{Mb/s}$ data bit rate takes about $\SI{118}{\mu s}$,
and duration shrinks to $\SI{84}{\mu s}$ if the nominal speed is increased to $\SI{1}{Mb/s}$.
Moving to \IOC\ results in $\SI{104}{\mu s}$ and $\SI{70}{\mu s}$ for the two nominal bit rates, respectively, 
with an increase in the net throughput over \EOC\ equal to $14\%$ and $20\%$.
As can be seen, 
reducing overheads could make CAN XL faster than $\SI{10}{Mb/s}$ Ethernet for these IP communications, and helps preventing congestion when sustained traffic comes from Ethernet.

\begin{figure*}[htbp]
    \centerline{\includegraphics[width=0.89\linewidth]{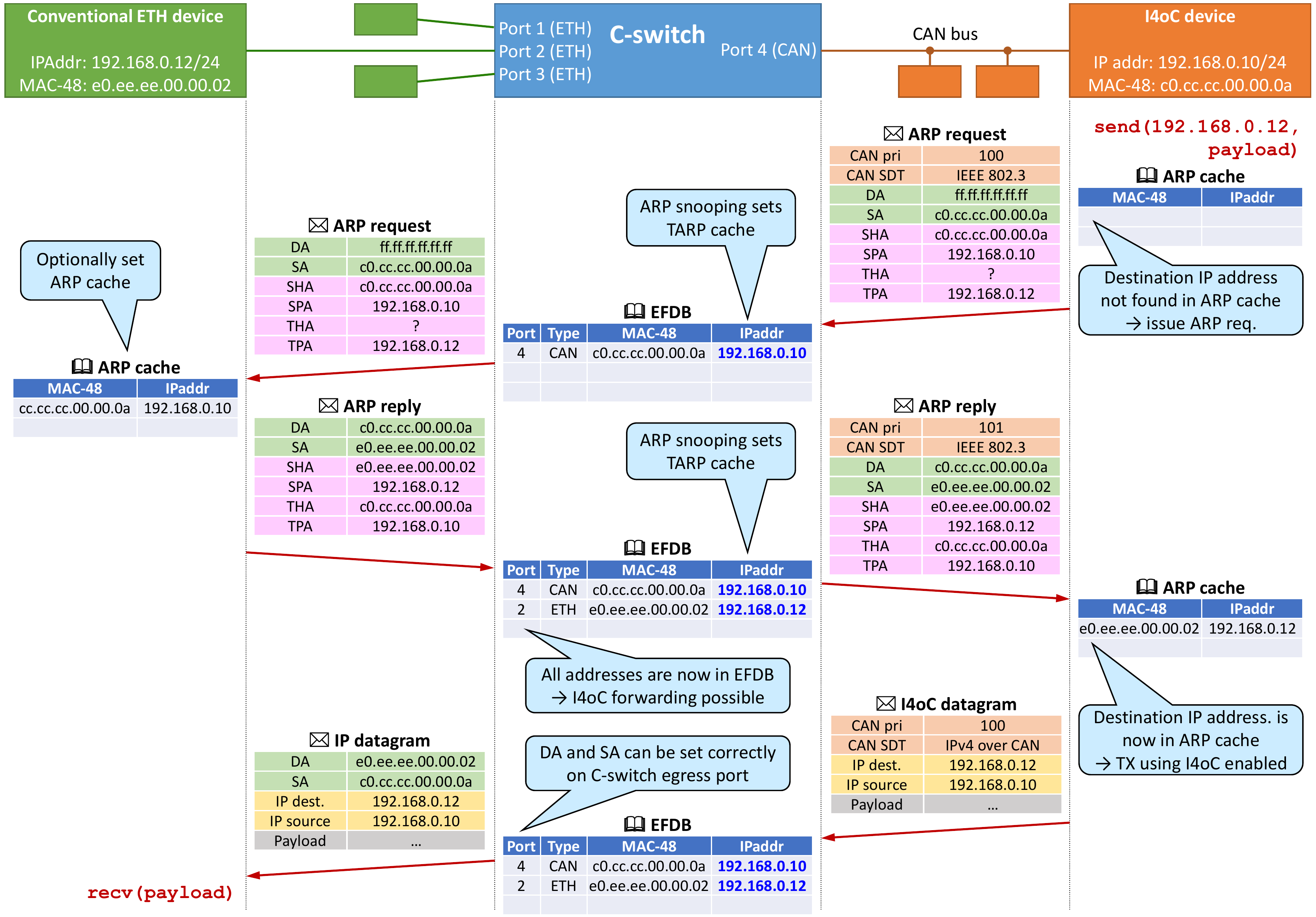}}
    \vspace{-0.1cm}
    \caption{Sample CAN-to-Ethernet datagram transmission using \IOC\ 
    (learning takes place in the EFDB through ARP snooping).}
    \label{fig:CAN1}
    \vspace{-0.1cm}
\end{figure*}

\subsection{\IOC\ Operation}
\IOC\ jointly manages aspects that belong to the data-link and network layers, 
hence it resembles Ethernet under several respects.
When the target node(s) are connected to the same CAN bus as the source, 
\IOC\ datagrams are delivered directly by exploiting broadcast transmissions on the bus.
Storing the whole destination IP address in AF enables receivers to quickly filter out irrelevant datagrams.
In the IPv6 case (not dealt with here), a suitable mapping function must be defined that, e.g., calculates a $\SI{4}{B}$ hash of the $\SI{16}{B}$ destination IP address, with the goal to avoid clashes.
Instead, when source and destination are attached to different network segments, datagrams are relayed by \mbox{C-switches}.
On the whole, the embedded network we are considering coincides with a single logical IP subnetwork,
and any destination can be reached directly with no hops across routers:
\mbox{C-switches} operate at the data-link layer and do not take part to the routing operations performed by IP.

The idea is to extend selective forwarding so that it operates with \IOC\ frames as well.
Dealing with such frames (that do not include MAC-48 DA and SA) requires that an extended filtering database (EFDB) is employed, which stores also the IP address of nodes.
This information is obtained by analyzing ARP messages.
For example, when an Ehernet/\EOC\ frame that embeds a ARP request is heard, the sender protocol address (SPA) field is used to create an entry $\langle$SA,SPA,port$\rangle$ in the EFDB for that port.
This resembles ARP snooping \cite{ARPsnoop}, but it is not related to security.
This approach is not more vulnerable against ARP spoofing attacks than switched Ethernet:
in fact, compromised nodes may impersonate other nodes by changing their MAC-48 addresses, 
hence corrupting the FDB also in conventional switches.
The above mechanism, we denote Transparent ARP (TARP), ``moves'' the information in the ARP table of \IOC\ nodes to \mbox{C-switches}, and loosely resembles Proxy ARP.
Practically, the EFDB can be implemented by complementing the conventional FDB in EWS with a separate data structure we call TARP cache, whose entries map IP addresses onto MAC-48 ones.
TARP learning can additionally rely on non-\IOC\ datagrams: when they are received by the \mbox{C-switch}, the source MAC-48 and IP addresses are cached.

Upon arrival of a datagram to the \mbox{C-switch}, the EFDB is inspected for a match.
If reception occurred on Ethernet or as \EOC, a MAC-48 DA is available, and a matching entry is looked for in the EFDB.
Instead, the destination of \IOC\ datagrams is searched among IP addresses.
As in conventional switches, if the destination IP address of an \IOC\ datagram is unknown, flooding takes place by forwarding it on all its ports.
However, this is unlikely to happen, as transmission of datagrams in end-nodes provided with a conventional IP protocol stack is preceded by a lookup to the local ARP table.
On the first query to obtain the target hardware address (THA),
an ARP request is broadcast by the sender, to which the target reacts with a unicast ARP reply.
These ARP messages enable backward learning in both directions on all the traversed switches and \mbox{C-switches},
and set the FDB/EFDB entries for the two end-nodes involved in the communication.
As soon as the \mbox{C-switch} understands on which port a certain \IOC\ node (characterized by its IP address) is connected,
selective forwarding is enabled, hence achieving traffic confinement.

If the source and destination nodes are attached to two distinct CAN buses connected to the same \mbox{C-switch},
or the path between them consists of a sequence of CAN buses interconnected by \mbox{C-switches},
the above mechanism is sufficient to support \IOC\ communication.
The more general case where some portion of the path consists of Ethernet links or uses \EOC\ requires specific mechanisms.
Forwarding datagrams from Ethernet (or \EOC) to \IOC\ is straightforward, and simply consists of using the streamlined header
of the latter.
The reverse case, where an \IOC\ datagram has to be forwarded, is a bit more complex.
In fact, its MAC-48 DA and SA are unavailable, and must be determined in some way before the datagram can be relayed, embedded into a well-formed Ethernet frame (possibly encapsulated as \EOC).

ARP snooping permits all the traversed \mbox{C-switches} (one or more) to learn the IP and MAC-48 addresses of both communicating end-points contextually.
Since EFDBs are initialized before \IOC\ datagrams are exchanged, outgoing Ethernet frames can be correctly reconstructed for them.
This is shown in the diagram of Fig.~\ref{fig:CAN1}, which shows that the \IOC\ transmission at the bottom of the figure succeeds to find the required information in the relevant EFDB entries
of the traversed \mbox{C-switch}.

Unfortunately, this does not work for static entries in the ARP table of end-nodes.
In this case, no ARP request-reply transactions are performed prior to the datagram exchange.
A possible solution is to mandate nodes to broadcast a gratuitous ARP reply before any IP datagrams are sent, for instance at startup, similarly to what is done by the IPv4 Address Conflict Detection (ACD) mechanism (RFC5227).
This enables all \mbox{C-switches} to properly set the entry related to every node in their EFDB using ARP snooping.
To improve robustness, end-nodes could be forced to send, from time to time (e.g., upon a suitable timeout expiration), 
one datagram as \EOC\ instead of \IOC, so as to fully refresh their EFDB entry in \mbox{C-switches}.

The decision on whether to use \EOC\ (default) or \IOC\ (optional) should be left to the system designer, 
possibly on a port-by-port basis when configuring end-nodes and \mbox{C-switches} 
(e.g., to optimize those CAN buses where utilization is particularly high).
The above mechanism for managing \EOC/\IOC\ transmissions in composite CAN XL-Ethernet networks is denoted uniform network addressing and traversal (UNAT).

\section{Bridges for Legacy CAN}
\label{sec:CANbridge}
One of the key requirements of automotive and embedded systems is to keep costs as low as possible.
This implies that unnecessary network equipment has to be avoided.
In composite CAN XL-Ethernet systems, at least one \mbox{C-switch} is needed.
A reasonable design goal is that, possibly, there is just a single such device onboard, which takes care of additional functions
besides interconnecting \EOC/\IOC\ CAN XL nodes and conventional Ethernet devices.

A useful function that could be easily carried out by the \mbox{C-switch} is to decouple data exchanges on distinct legacy CAN buses,
each of which constitutes a separate contention (arbitration) domain, analogous to the collision domain in legacy CSMA/CD Ethernet.
Bridges, which operate according to the store and forward principle,
are often employed in real CAN networks
to enlarge their extension and increase the overall bandwidth \cite{RTSS20202}.
Unlike Ethernet bridges, they are not provided with backward learning and do not perform flooding, 
in such a way to strictly confine local traffic.
Instead, they are statically configured to relay specific CAN frames, each one characterized by its identifier, 
from one port to another (or several others).
In doing so, the original identifier field is possibly changed to prevent clashes on the target bus, yet preserving some guarantees about timeliness.
This is because the number of available CAN IDs is quite small ($2048$ for base $\SI{11}{b}$ identifiers)
and does not permit global assignment, so that they typically have local validity.
Similarly, CAN frames can be also forwarded to/from Ethernet ports.

Unfortunately, to the best of our knowledge no widely agreed specifications exist today about the way 
such forwarding operations can be generically described and implemented in a standard way.
Conversely, many effective, but incompatible proposals and commercial products are available.
Therefore, a flexible, plug\&play solution is sought that can be readily embedded in any new designs.
It is worth pointing out that 
the role of above CAN bridges can be carried out by \mbox{C-switches},
which become pivotal to the controlled relay of every kind of traffic (Ethernet, IP, and CAN) among different segment/busses in composite networks.
Defining and classifying agreed forwarding rules and techniques for the different protocols 
(e.g., CAN to Ethernet/UDP/TCP, and vice-versa) is part of our future work.

\section{Conclusions}
\label{sec:Conc}
CAN XL has been conceived to overcome the limitations of legacy CAN, preserving at the same time a very good degree of compatibility with it.
Although CAN XL speed can easily exceed $\SI{10}{Mb/s}$, it is still noticeably lower than Fast and Gigabit Ethernet.
For this reason, great care has been taken in its definition so that the two kinds of solutions may coexist with limited efforts in the same system, thanks to encapsulation techniques supported by the SDU type.

Central to the seamless operations in a composite CAN XL-Ethernet network is the \mbox{C-switch},
which resembles conventional Ethernet switches but is provided with ports of both \TETH\ and \TCAN\ type.
It permits frames to be transparently forwarded between devices of the two kinds.
In this paper two solutions are described: in the former, we term \EOC, Ethernet frames are suitably embedded in CAN XL frames,
which is what CAN XL designers intended.
It is a general-purpose approach that privileges connectivity.
The latter, we call \IOC, is instead useful when applications communicate using the TCP/UDP/IP stack, 
which is what customarily happens in the real world.
In this case, communication efficiency can be improved up to $20\%$ for specific kinds of traffic, characterized by small IP datagrams, adopting streamlined protocol headers.
Compatibility with \EOC\ is ensured by means of extended filtering databases, which also store information about IP addresses.
Minimal modifications are required to the protocol stack, which performs learning using ARP snooping techniques.

Unlike gateways, which have been widely used in the past years in automotive, embedded, and industrial scenarios as a custom approach to support interconnection of heterogeneous devices,
\mbox{C-switches} are meant to offer a standard solution for composite CAN XL-Ethernet networks.
On the one hand this reduces design and implementation cost, on the other they may set a universal and agreed way to achieve full interoperability between the two solutions in the long term.

\bibliographystyle{IEEEtran}
\bibliography{mybib}

\cleardoublepage

\end{document}